# NEAT, An Astrometric Telescope To Probe Planetary Systems Down To The Earth Mass Around Nearby Solar-Type Stars

F. Malbet[1], A. Léger[2], R. Goullioud[3], M. Shao[3], P.-O. Lagage[4], C. Cara[4], G. Durand[4], P. Feautrier[1], B. Jakobsson[5], E. Hinglais[6], M. Mercier[7] and the NEAT collaboration[8]

[1]UJF-Grenoble 1 / CNRS-INSU, Institut de Planétologie et d'Astrophysique de Grenoble (IPAG), UMR 5274, BP 53, F-38041 Grenoble cedex 9, France; [2]Université Paris Sud CNRS-INSU, Institut d'Astrophysique Spatiale (IAS) UMR 8617, Bât 120-121, F-91405 Orsay cedex, France; [3]Jet Propulsion Laboratory (JPL), California Institute of Technology, Pasadena CA 91109, USA; [4]Laboratoire AIM, CEA-IRFU / CNRS-INSU / Université Paris Diderot, CEA Saclay, France; [5]Swedish Space Corporation (SSC), Solna, Sweden; [6]Centre National d'Etudes Spatiales (CNES), Toulouse, France; [7]Thales Alenia Space (TAS), Cannes, France; [8]Full list of NEAT proposal members at
http://neat.obs.ujf-grenoble.fr

## Abstract

The NEAT (Nearby Earth Astrometric Telescope) mission is a proposition submitted to ESA for its 2010 call for M-size mission within the Cosmic Vision 2015-2025 plan [1]. The main scientific goal of the NEAT mission is to detect and characterize planetary systems in an exhaustive way down to 1 Earth mass in the habitable zone and further away, around nearby stars for F, G, and K spectral types. This survey would provide the actual planetary masses, the full characterization of the orbits including their inclination, for all the components of the planetary system down to that mass limit. Only extremely-high-precision astrometry, in space, can detect the dynamical effect due to even low mass orbiting planets on their central star, reaching those scientific goals. NEAT will continue the work performed by Hipparcos (1mas precision) and Gaia (7μas aimed) by reaching a precision that is improved by two orders of magnitude (0.05μas, 1σ accuracy). The NEAT mission profile is driven by the fact that the two main modules of the payload, the telescope and the focal plane, must be placed 40m away leading to a formation flying option that has been studied as the reference mission. The NEAT satellites are foreseen to operate at L2 for 5 years, the telescope satellite moving around the focal plane one to point different targets and allowing whole sky coverage in less than 20 days. The payload is made of 3 subsystems: primary mirror and its dynamic support, the focal plane with the detectors, and the metrology. The principle is to measure the angles between the target star, usually bright (R ≤ 6), and fainter reference stars (R ≤ 11) using a metrology system that projects dynamical Young's fringes onto the focal plane. The proposed mission architecture relies on the use of two satellites of about 700 kg each in formation flying, offering a capability of more than 20,000 reconfigurations. The two satellites are launched in a stacked configuration using a Soyuz ST launch, and are deployed after launch in order to individually perform cruise to their operational Lissajous orbit.

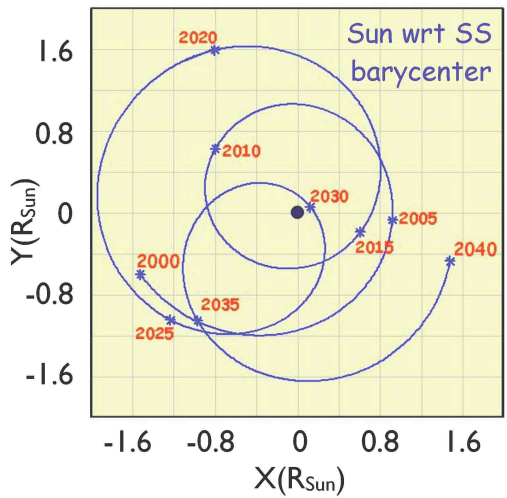 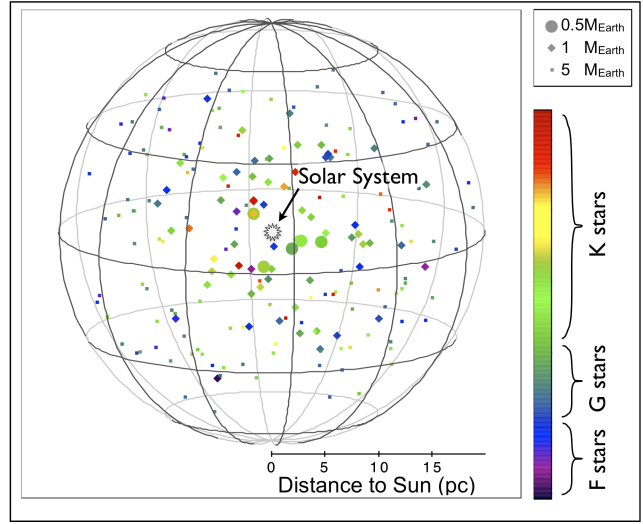

**Figure 1** *Astrometric motion of the Sun due to the presence of the planets of the Solar System.*

**Figure 2** *NEAT targets around the Sun. They correspond to a volume-limited sample of all stars with spectral type between F and K. The sizes of the symbols correspond to the detection limit indicated in the upper right.*

## 1. Science Objectives

The prime goal of NEAT is to detect and characterize planetary systems orbiting bright stars in the solar neighborhood that have a planetary architecture like that of our Solar System or an alternative planetary system made of Earth mass planets.

The principle for the detection of the planetary systems is to measure the reflex motion of the star due to the presence of the different planetary components (since the star orbits around the center of mass of the systems composed of the star and its planets). The reflex motion of the Sun because of the presence of Jupiter, Saturn, and the other planets of the Solar System is plotted on Figure 1 as an example of what can be expected. At 10 pc the motion due the presence of Jupiter is about 500 µas, whereas the amplitude due to the presence of the Earth at 1AU is 0.3 µas! The detection limit depends linearly on the distance between the star and the planet, the mass of the star, the mass of the planet and the distance of the system to the observer. NEAT aims at measuring motions down to a floor value of 0.05 µas, allowing detection of Earth-mass planets orbiting at 1 AU around Sun-mass stars located 10 pc away with a SNR of 6. The targets plotted in Figure 2 correspond to an exhaustive search of 1 Earth mass planets (resp. 5 Earth mass planets) around K stars located at distances from the Sun up to 6 pc (resp. 12 pc), G stars up to 10 pc (resp. 17 pc), and F stars up to 14 pc (resp. 19 pc), in the whole HZ of the star, excluding spectroscopic binaries and very active stars.

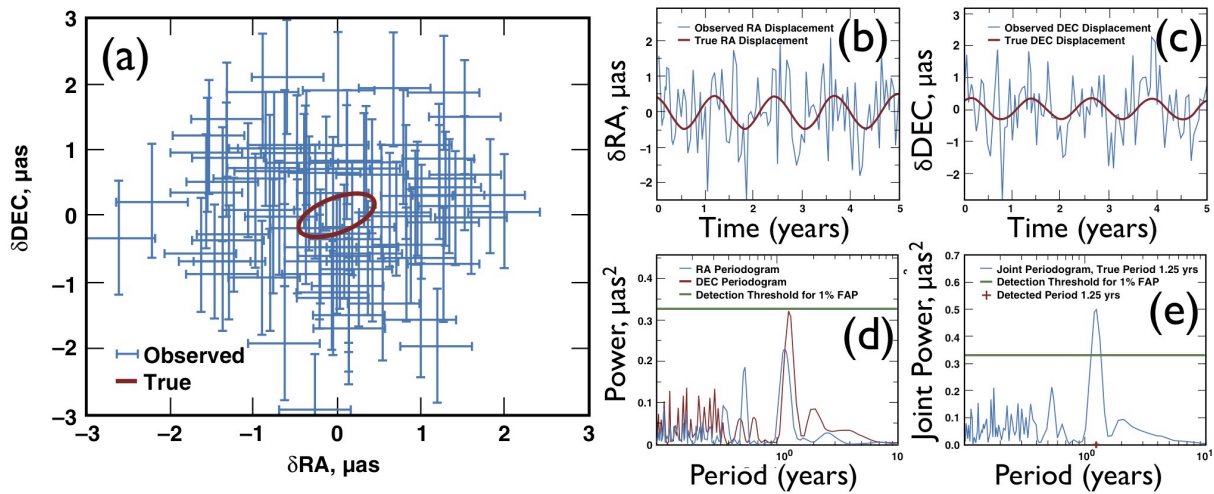

*Figure 3* Simulation of the astrometric detection of a planet (1.5 Earth mass orbiting a 1 solar mass star at 1.16 AU, located at 10 pc) with a SNR=6. (a) Sky plot showing the astrometric orbit (solid brown curve) and the NEAT measurements with error bars (in blue); (b) and (c) same data but shown as time series of the right ascension and declination astrometric signal; (d) Separated periodograms of right ascension (blue line) and declination (brown line) measurements. (e) Joint periodogram from R.A. and DEC simultaneously. Whereas the orbit cannot be determined from the astrometric signal without the time information, its period is reliably detected in the joint periodogram (1.25yr), with a false-alarm probability below 1% (green line). Then, the planetary mass and orbit parameters can be determined by fitting the astrometric measurements.

Figure 3 shows the type of orbit that can be obtained with an Earth-mass planet located at 10 pc. NEAT observations will allow the detection around nearby stars of planets equivalent to Venus, Earth, (Mars), Jupiter, and Saturn, with orbits possibly similar to those in our Solar System. The NEAT mission will answer the following questions:

- What are the dynamical interactions between giant and telluric planets in a large variety of systems?

- What are the detailed processes involved in planet formation as revealed by their present configuration?

- What are the distributions of architectures of planetary systems in our neighborhood up to D ≈ 15pc?

- What are the masses, and orbital parameters, of telluric planets that are candidates for future direct detection and spectroscopic characterization missions?

Special emphasis will be put on planets in the Habitable Zone (HZ) because this is a region of prime interest for astrobiology. Indeed orbital parameters obtained with NEAT will allow spectroscopic follow-up observations to be scheduled precisely when the configuration is the most favorable.

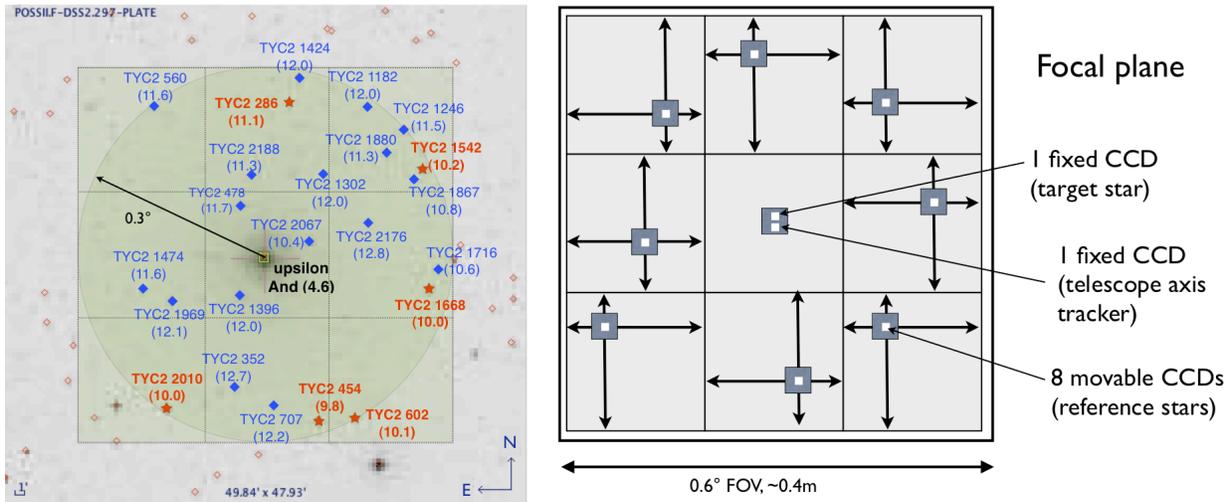

**Figure 4** Left: a typical field of view around a nearby solar-type like Upsilon Andromae. The target star is bright (R≤6) and 6 to 8 stars can be found within 0.3° brighter than (R≤11) to serve as reference stars. Right: The focal plane of NEAT (0.4mx0.4m) hosts 10 CCDs. The 8 peripheral ones can be moved to be centered on each reference star, while the 2 central ones are used to stabilized the field on the target stars.

## 2. NEAT Instrumental Concept

Left part of Figure 4 shows a typical field around NEAT targets with the presence of about ten reference stars. With typical 1 m telescope and 10 µm-sized detector pixels, one needs 30,000 x 30,000-pixel detectors or alternatively, or ten 512 x 512 CCDs, from which 8 are mobiles in order to cover all possible reference stars (right part of same figure). To measure the angle between the target star and the several reference stars, one introduce an interferometric metrology calibration system. The idea is to project laser beams from the primary mirror to produce Young's fringes onto the focal plane that are used to calibrate the distances.

The NEAT concept (Figure 5) is based on a long focal-length telescope with a single parabolic mirror rather than on a classical three-mirror anastigmat (TMA) design to avoid any beam walk errors, but also to because the unavailability of suitable metrology and stability of biases and robustness against proper motion that would kill any measurements at the required level of precision.

The size of the NEAT mission could be reduced (or increased) with a direct impact on the accessible number of targets but not in an abrupt way. For instance, for a same amount of integration time and number of maneuvers, the options listed in Table 1 are possible, with impacts on the number of stars that can be investigated down to 0.5 and 1 Earth mass. The time necessary to achieve a given precision depends on the mass limit that we want to reach: going from 0.5 M⊕ to 1 M⊕ requires twice less precision and therefore 4 times less observing time. This is even stronger to go from 1 M⊕ to 5 M⊕ with a reduction of observing time by target of 25. The observing time is driven by the most demanding precision and therefore, we can keep a 200 target list for the total

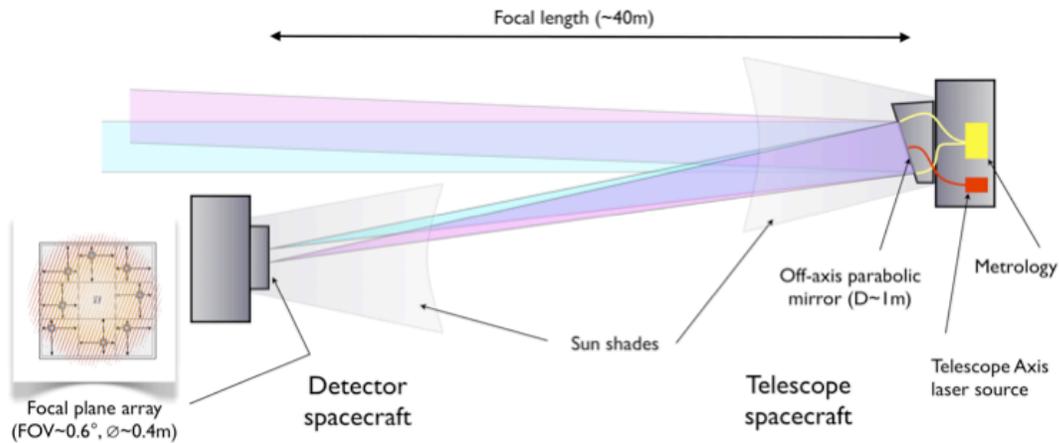

*Figure 5* NEAT instrumental concept with two satellites: the telescope satellite (right) that carries the parabolic primary 1m mirror and the source for the metrology, and the detector satellite which carries the focal plane that is illuminated once every minute by Young's fringes to calibrate the position of all CCDs.

*Table 1:* Science impact from NEAT scaling

| Mission name | Mirror diameter | Focal length | Field of view diameter | Focal Plane size | DMA in 1h | # targets for a given mass limit | | |
|---|---|---|---|---|---|---|---|---|
| | (m) | (m) | (deg) | (cm) | (µas) | 0.5M⊕ | 1 M⊕ | 5 M⊕ |
| NEAT plus | 1.2 | 50 | 0.45 | 40 | 0.7 | 7 | 100 | 200 |
| NEAT | 1.0 | 40 | 0.56 | 40 | 0.8 | 5 | 70 | 200 |
| NEAT light | 0.8 | 30 | 0.71 | 35 | 1.0 | 4 | 50 | 200 |
| EXAM | 0.6 | 20 | 0.85 | 30 | 1.4 | 2 | 35 | 200 |

DMA = Differential astrometric Measurement Accuracy (rms)

number of targets and adjust the number of systems that will be investigated at 1 M⊕ and 0.5 M⊕ in Table 1. There is room for adjustment keeping in mind that one wants to survey the neighborhood with the smallest mass limit possible.

## 3. NEAT Spacecraft

As explained in Section 2, in order to achieve sub-µas accuracy, one needs to have a single mirror telescope design with a primary aperture that focus the signal on the detector plane. Furthermore to reach the spatial resolution of the telescope at visible wavelength and the field of view necessary to find relatively bright reference stars with limited optical aberration, we need a focal length in the order of tens of meters with a 1m-class telescope. Discarding the TMA option, two options are left to accommodate the long focus telescope: a single spacecraft with a deployable boom or two spacecrafts flying in formation. The single spacecraft solution is better adapted to rather short focal length, while the double spacecraft solution can accommodate long focal length. With the NEAT design with a 1-m telescope and 40-m focal length, we felt that the formation flying was more appropriate but the single spacecraft has indeed been studied in a JPL

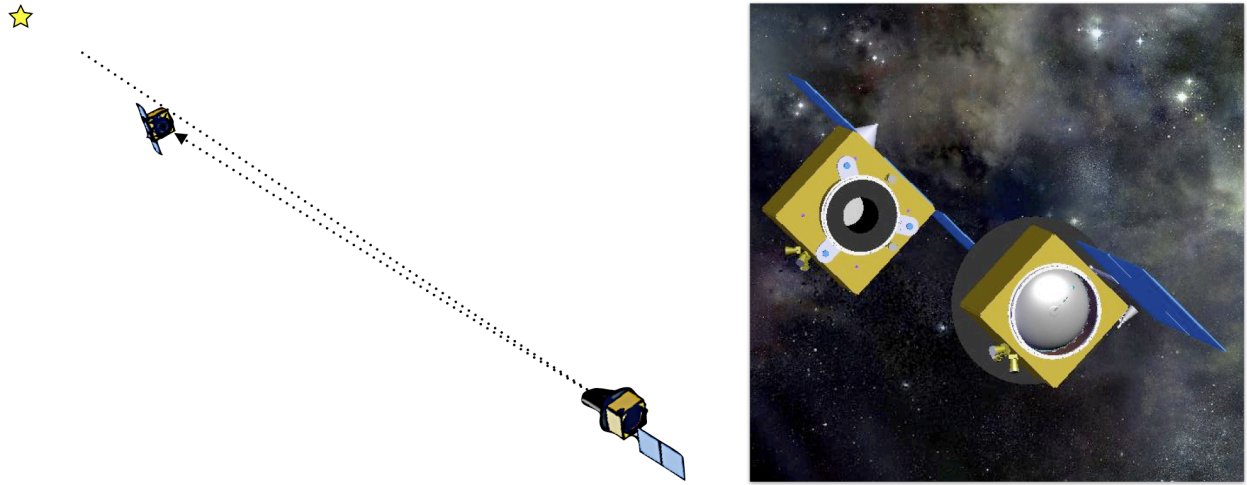

*Figure 6* Left: NEAT spacecraft in operation with the two satellites separated by 40 m. Right: closer external view of the two satellites.

smaller version of NEAT called EXAM (Exoplanet Astrometric Mission).

Figure 6 shows NEAT spacecraft in operation. The telescope satellite is observing the star while the focal plane satellite is located 40m away not far from the line of sight is looking back toward the telescope satellite.

### 3.1. System Functional Description

The formation flying has to ensure anti-collision and safeguarding of the flight configuration, based on the successful PRISMA flight heritage. In addition, the spacecraft typically performs 12 to 20 daily reconfigurations of less than 10° of the system line of sight corresponding to 7m of translation of one satellite compared to the other perpendicular to the line of sight. During these configurations, the telescope satellite performs translations —supported by the FFRF Units— using its large hydrazine tanks (250 kg) for a $\Delta V \sim 605$ m/s. When the two satellites approach the required configuration, the telescope satellite freezes, and the focal plane satellite performs fine relative pointing control using micro-propulsion system to ensure a longitudinal and lateral position accuracy of the 2 spacecraft of the order of 2 mm. As a result, the µ-propulsion has to compensate for hydrazine control inaccuracies, which requires large nitrogen gas tanks (92 kg for $\Delta V \sim 75$ m/s). Finally, 28 kg of hydrazine in the FP satellite allows $\Delta V \sim 55$ m/s for station keeping and other operations.

### 3.2. Satellite Design Description

The design of the two satellites is based on an 1194 mm central tube architecture, which allows a low structural index for the stacked configuration and provides accommodation for payloads and large hydrazine tanks. Strong heritage does exist on the two satellites avionics and AOCS. In addition, they both require similar function, which would allow introducing synergies between the two satellites for design, procurement, assembly, integration and tests. The proposed AOCS configuration is a gyroless architecture relying on reaction wheels and high-performance star trackers (Hydra Sodern), which is compatible with a 3 arcsec pointing accuracy. The satellites communication subsystems use X-Band active pointing antenna, supported by large gain antenna for low Earth orbit positioning and cruise, coupled with a 50 W RF Transmitter. The active pointing medium gain antenna allows simultaneous data acquisition and downlink. A reference solution for the satellite on-board computer could rely on the Herschel-Planck avionics.

The two satellites would have custom mechanical-thermal-propulsion architectures. The telescope satellite features a dry mass of 724 kg and the focal plane satellite a dry mass of 656 kg. The focal plane satellite carries the stacked configuration. The payload (focal plane + baffle) is assembled inside the central tube, which also ensures the stacked configuration structural stiffness. The spacecraft bus, and large cold gas tanks, is assembled on a structural box carried by the central tube. The proposed architecture uses a large hydrazine tank inside the central tube, which offers a capacity of up to 600 kg hydrazine, thus allowing both a low filling ratio and a large mission growth potential. The payload module —with the payload mirror, rotating mechanisms and baffle— is then assembled on the central tube.

*3.3. Proposed Procurement Approach*

The NEAT mission is particularly adapted to offer a modular spacecraft approach, with simple interfaces between payload and spacecraft bus elements. For both satellites, the payload module is clearly identified and assembled inside the structural central tube. In addition, a large number of satellite building blocks can be common to the two satellites, in order to ease mission procurement and tests. This configuration is particularly compatible with the ESA procurement scheme. The payload is made of 3 subsystems: primary mirror and its dynamic support, the focal plane with its detectors and the metrology.

## 4. Challenges

Several challenges must be taken into account both on the astrophysical side and the technical one:

- **Stellar activity**. Spots and bright structures on the stellar surface induce astrometric, photometric and RV signals. Using the Sun as a proxy Lagrange et al. [2] have computed the astrometric, photometric and RV variations that would be measured from an observer located 10 pc away. It appears that the astrometric variations due to spots and bright structures are small compared to the signal of an Earth mass planet in the HZ. In our target list, we keep only stars for which their intrinsic activity should not prevent the detection of an Earth-mass planet, even during its high activity period. This corresponds to 96% of the initial star list.

- **Perturbations from reference stars**. Because the reference stars are much more distant (≈ 1 kpc) than the target star (≈ 10 pc), NEAT is 100 times less sensitive to their planetary perturbations. Only Saturn-Jupiter mass objects matter, and statistically, they are only present around ≈ 10% of stars. These massive planets can be searched for by fitting first the reference star system, possibly eliminate those with giant planets, and studying the target star with respect to that new reference frame.

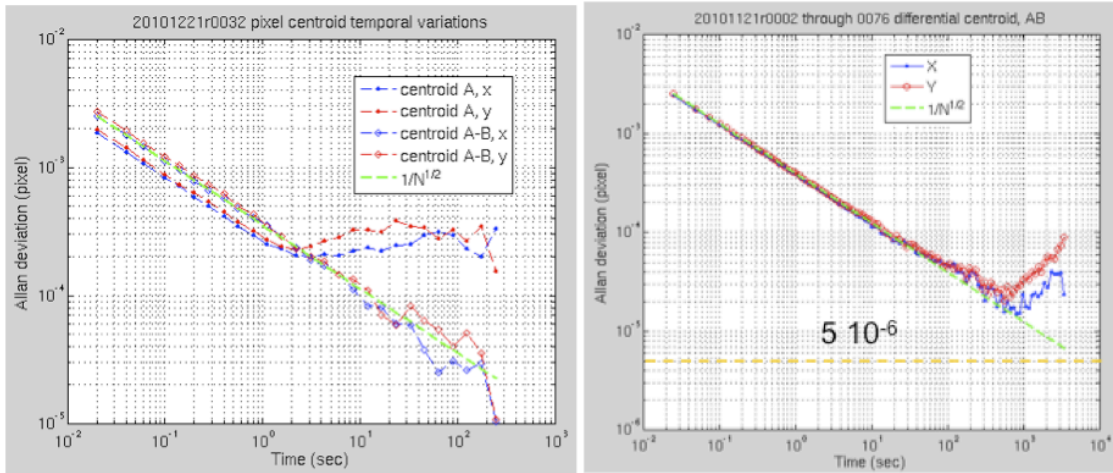

**Figure 7** *Latest results obtained from the CCD/metrology test bench.*

- **Radial velocity (RV) screening** . To solve unambiguously for giant planets with periods longer than 5 yrs, it is necessary to have a ground RV survey for 15 yrs of the 200-selected target stars, at the presently available accuracy of 1m/s. RV is already observing more than 80% of our targets, but the observations of the rest of them should start soon, well before the whole NEAT data is available.

- **Flexibility of objectives to upgrades / downgrades of the mission**. One of the strengths of NEAT is its flexibility, the possibility to adjust the size of the instrument with impacts on the science that are not prohibitive. The size of the NEAT mission could be reduced (or increased) with a direct impact on the accessible number of targets but not in an abrupt way (see Table 1).

- **Optical aberrations**. NEAT uses a very simple telescope optical design. A 1-m diameter clear aperture off-axis parabola, with an off- axis distance of 1 m and a 40 m focal length. The focal plane is at the prime focus. The telescope is diffraction limited at the center of the field, where the target stars will be observed, but coma produces some field dependent aberrations. However the impact remains low since we are looking at differential effects.

- **Centroid measurements.** Inter-pixel response variations exist in real CCDs, which we calibrate by measuring the pixel response of each pixel in Fourier space. Capturing inter-pixel variations of pixel response to the third order terms in the power series expansion, we have shown with simulated data that the centroid displacement estimation is accurate to a few micro-pixels [3].

- **Stability of the primary mirror.** While the wavefront deviations to optimal shape causes a centroid shift of ≈ 6 − 10 µas ($10^{-4}$ pixels), differential errors remains less than ≈ 0.3 µas ($3 \times 10^{-6}$ pixels).

- **CCD damage in L2 environment.** CCDs, like most semiconductors, suffer damage in radiation environments such as encountered by space missions. One particular performance parameter, Charge Transfer Efficiency (CTE), degrades with known consequences on the efficiency of science missions like Gaia. However there are a number of important differences between NEAT and Gaia, which justify the assumption that radiation damage effects will play a much smaller role.

- **CCD/metrology tests in the lab.** In the absence of optical errors, the major error sources are associated with the focal plane: (1) motions of the CCD pixels, which have to be monitored to $3 \times 10^{-6}$ pixels every 60 s, i.e. 0.03 nm; (2) measurements of the centroid of the star images with $5 \times 10^{-6}$ pixel accuracy. We have set up technology testbeds to demonstrate that we can achieve these objectives (Figure 7). The technology objective for (1) has almost been reached and the technology demonstration for (2) is underway and should be completed soon.

## 5. Perspectives

In the Cosmic Vision plan for 2015-2025, the community has identified in Theme 1 the question: "*What are the conditions for planet formation?*" and the recommendation in Sect. 1.2: "*Search for planets around stars other than the Sun...*" ultra high precise astrometry as a key technique to explore our solar-like neighbors.

> *"On a longer timescale, a complete census of all Earth-sized planets within 100 pc of the Sun would be highly desirable. Building on Gaia's expected contribution on larger planets, this could be achieved with a high-precision terrestrial planet astrometric surveyor."*

We have designed NEAT to be this astrometric surveyor. We believe that there is a place for a mission like NEAT in future space programs, that is to say, a mission that is capable of detecting and characterizing planetary systems orbiting bright stars in the solar neighborhood that have a planetary architecture like that of our Solar System or an alternative planetary system partly com- posed of Earth-mass planets. These stars, visible with the naked eye or simple binoculars and if found to host Earth- mass planets, will change humanity's view of the night sky.

## Acknowledgement

This work has benefited support from Centre National des Etudes Spatiales (CNES), Jet Propulsion Laboratory (JPL), Thales Alenia Space (TAS) and Swedish Space Corporation (SSC).